\documentstyle[12pt]{article}
\begin{document}

\begin{center}
{\bf GEODESICS IN LEWIS SPACETIME}
\end{center}

\vspace*{.5cm}

\centerline{{\bf L. Herrera}\footnote{On leave from Departamento de
F\'{\i}sica, Facultad de Ciencias, Universidad Central de Venezuela,
Caracas, Venezuela and Centro de Astrof\'{\i}sica Te\'orica, Merida,
Venezuela.}}
\begin{center}
Area de F\'{\i}sica Te\'orica, Grupo de Relatividad, Universidad de
Salamanca, Salamanca, Spain.
\end{center}

\centerline{\bf N. O. Santos}
\begin{center}
Universit\'e Paris VI, CNRS/URA 769, Cosmologie et Gravitation
Relativistes, Tour 22-12, 4\'eme \'etage, Bo\^{\i}te 142, 4 place Jussieu,
75005 Paris, France.
\end{center}
\begin{center}
Observat\'orio Nacional CNPq, Departamento de Astrof\'{\i}sica, rua General
Jos\'e Cristino 77, 20921 Rio de Janeiro, Brazil.
\end{center}

\begin{abstract}
The geodesic equations are integrated for the Lewis metric and the effects
of the different parameters appearing in the Weyl class on the motion of
test particles are brought out. Particular attention deserves the
appearance of a force parallel to the axial axis and without Newtonian
analogue.
\end{abstract}


\newpage

\section{Introduction}
Lewis metric \cite{Lewis} describes the spacetime outside an axially
symmetric unbounded, along the axial axis, source, endowed with angular
momentum. Usually this metric is presented with four parameters
\cite{Kramer} which may be real (Weyl class) or complex (Lewis class).
In recent papers \cite{Silva,Silva1}, the physical meaning of these
parameters have been discussed for both classes. Thus for the Weyl class,
it appears that one of the parameters is proportional to the energy per
unit length (at least in the Newtonian limit), a second parameter is the
arbitrary constant potential which is always present in the Newtonian
solution, and the remaining two parameters are responsible for the non
staticity of the metric, although affecting staticity in different ways.

It is the purpose of this work to delve more deeply into the physical
meaning of the aforesaid parameters. Accordingly we shall study the motion
of test particles in the Lewis spacetime. In the next section we shall
present the Lewis metric, the geodesic equations are given in section 3 and
the specific case of circular geodesics is considered in section 4. The two
next sections are devoted to the Weyl class. Two specific examples for
circular geodesics are studied in section 5 exhibiting the difference in
the role played by the two parameters responsible for the non staticity.
Radial geodesics are considered for several different cases in the last
section. Furthermore, in this last section, the geodesics along the axial
axis shows the appearance of a force parallel to the axis, without
Newtonian analogue. This force tends to damp the motion along the $z$ axis,
when the particle approaches this axis, whereas this tendency reverses when
the particle moves away from it.

As it will be seen below, besides the usual dragging effect, to be expected
whenever a rotating source is involved, another, {\it mixed}, effect
appears, which we call {\it topological frame dragging}, related
simultaneously to the rotation of the source and to the topological deffect
associated with some metric parameters.

\section{Lewis spacetime}
The general line element for a cylindrically symmetric stationary spacetime
is given by
\begin{equation}
ds^2=-fdt^2+2kdtd\phi+e^{\mu}(dr^2+dz^2)+ld\phi^2,
\end{equation}
where $f,k,\mu$ and $l$ are functions only of $r$, and the ranges of the
coordinates $t,z$ and $\phi$ are
\begin{equation}
-\infty<t<\infty,\qquad -\infty<z<\infty,\qquad 0\leq\phi\leq 2\pi,
\end{equation}
the hypersurfaces $\phi=0$ and $\phi=2\pi$ being identified. The
coordinates are numbered
\begin{equation}
x^0=t,\qquad x^1=r,\qquad x^2=z,\qquad x^3=\phi.
\end{equation}
Einstein's field equations for vacuum are
\begin{equation}
R_{\mu\nu}=0,
\end{equation}
and with metric (1) the non identically null components become in van
Stockum's \cite{Stockum} notation,
\begin{eqnarray}
2e^{\mu}DR^0_0=\left(\frac{lf^{\prime}+kk^{\prime}}{D}\right)^{\prime},\\
2e^{\mu}DR^3_0=\left(\frac{fk^{\prime}-kf^{\prime}}{D}\right)^{\prime},\\
2e^{\mu}DR^0_3=\left(\frac{kl^{\prime}-lk^{\prime}}{D}\right)^{\prime},\\
2e^{\mu}DR^3_3=\left(\frac{fl^{\prime}+kk^{\prime}}{D}\right)^{\prime},\\
2R^1_1=-\mu^{\prime\prime}+\mu^{\prime}\frac{D^{\prime}}{D}
-2\frac{D^{\prime\prime}}{D}+\frac{k^{\prime 2}+f^{\prime}l^{\prime}}{D^2},\\
2R_{22}=-\mu^{\prime\prime}-\mu^{\prime}\frac{D^{\prime}}{D},
\end{eqnarray}
where the primes stand for differentiation with respect to $r$ and
\begin{equation}
D^2=k^2+fl.
\end{equation}
The four equations (5,6,7,8) are not all independent; any one of them can
be expressed in terms of the remaining three. The general solution of (4)
for (1) is the stationary Lewis metric \cite{Lewis} given in the notation
of \cite{Kramer}
\begin{eqnarray}
f=ar^{-n+1}-\frac{c^2}{n^2a}r^{n+1},\\
k=-Af,\\
e^{\mu}=r^{(n^2-1)/2},\\
l=\frac{r^2}{f}-A^2f,
\end{eqnarray}
with
\begin{equation}
A=\frac{cr^{n+1}}{naf}+b.
\end{equation}
The constants $n,a,b$ and $c$ can be either real or complex, and the
corresponding solutions belong to the Weyl or Lewis classes. For the Lewis
class these constants are given by
\begin{eqnarray}
n=im,\\
a=\frac{1}{2}(a^2_1-b^2_1)+ia_1b_1,\\
b=\frac{a_1a_2+b_1b_2}{a_1^2+b^2_1}+\frac{i}{a_1^2+b_1^2},\\
c=\frac{1}{2}m(a^2_1+b^2_1),
\end{eqnarray}
where $m,a_1,b_1,a_2$ and $b_2$ are real constants and satisfy
\begin{equation}
a_1b_2-a_2b_1=1.
\end{equation}
The equations (17,18,19,20,21) reveal that if the values of the parameters
$n$ and $a$, or $n$ and $b$ are known, we can obtain the parameter $c$.
However, knowing $n$ and $c$, we cannot obtain $a$ and $b$. The metric
coeficients (12,13,14,15) with (17,18,19,20) become \cite{Lewis}
\begin{eqnarray}
f=(a_1^2-b_1^2)r\cos(m\ln r)+2a_1b_1r\sin(m\ln r),\\
k=-(a_1a_2-b_1b_2)r\cos(m\ln r)-(a_1b_2+a_2b_1)r\sin(m\ln r),\\
e^{\mu}=r^{-(m^2+1)/2},\\
l=-(a_2^2-b_2^2)r\cos(m\ln r)-2a_2b_2r\sin(m\ln r).
\end{eqnarray}

\section{Geodesics}
The equations governing geodesics can be derived from the Lagrangian
\begin{equation}
2{\cal L}=g_{\mu\nu}\frac{dx^{\mu}}{d\lambda}\frac{dx^{\nu}}{d\lambda},
\end{equation}
where $\lambda$ is an affine parameter along the geodesics. For timelike
geodesics $\lambda$ is the proper time. From the extremal problem it
emerges the Euler-Lagrange equations
\begin{equation}
\frac{d}{d\lambda}\left(\frac{\partial{\cal L}}{\partial\dot x^{\alpha}}
\right)-\frac{\partial{\cal L}}{\partial x^{\alpha}}=0,
\end{equation}
and from them follow
the geodesics given by
\begin{equation}
\ddot x^{\alpha}+\Gamma^{\alpha}_{\beta\gamma}
\dot x^{\beta}\dot x^{\gamma}=0,
\end{equation}
where the overdot stands for differentiation with respect to $\lambda$.
For spacetime (1) the Lagrangian (26) is
\begin{equation}
2{\cal L}=-f\dot t^2+2k\dot t\dot \phi+e^{\mu}(\dot r^2+\dot z^2)+l
\dot\phi^2,
\end{equation}
and the geodesic equations (28) are
\begin{eqnarray}
D\ddot t+\frac{lf^{\prime}+kk^{\prime}}{D}\dot t\dot r+
\frac{kl^{\prime}-lk^{\prime}}{D}\dot r\dot\phi=0,\\
2\ddot r+e^{-\mu}(f^{\prime}\dot t^2-2k^{\prime}\dot t\dot\phi-l^{\prime}
\dot\phi^2)+\mu^{\prime}(\dot r^2-\dot z^2)=0,\\
\ddot z+\mu^{\prime}\dot r\dot z=0,\\
D\ddot\phi+\frac{fk^{\prime}-kf^{\prime}}{D}\dot f\dot r+\frac{fl^{\prime}
+kk^{\prime}}{D}\dot r\dot\phi=0.
\end{eqnarray}
The corresponding canonical momenta to (29) are
\begin{eqnarray}
p_t=-\frac{\partial{\cal L}}{\partial\dot t}=f\dot t-k\dot\phi,\\
p_r=\frac{\partial{\cal L}}{\partial\dot r}=e^{\mu}\dot r,\\
p_z=\frac{\partial{\cal L}}{\partial\dot z}=e^{\mu}\dot z,\\
p_{\phi}=\frac{\partial{\cal L}}{\partial\dot\phi}=k\dot t+l\dot\phi.
\end{eqnarray}
The integrals of motion follow from the Euler-Lagrange equations (27)
\begin{eqnarray}
\frac{dp_t}{d\lambda}=\frac{\partial{\cal L}}{\partial t}=0,\\
\frac{dp_z}{d\lambda}=\frac{\partial{\cal L}}{\partial z}=0,\\
\frac{dp_{\phi}}{d\lambda}=\frac{\partial{\cal L}}{\partial\phi}=0,
\end{eqnarray}
and the conserved quantities are
\begin{equation}
p_t=E,\qquad p_z=P,\qquad p_{\phi}=L,
\end{equation}
where the constants $E,P$ and $L$ represent respectively the total energy
of the test particle, its momentum along the $z$ axis and its angular
momentum about the $z$ axis.\begin{eqnarray}
\dot t=\frac{1}{D^2}(Lk+El),\\
\dot z=Pe^{-\mu},\\
\dot\phi=\frac{1}{D^2}(Lf-Ek),
\end{eqnarray}
and instead of integrating (31) we can use the line element (1) to obtain
\begin{equation}
-\epsilon=-f\dot t^2+2k\dot t\dot\phi+e^{\mu}(\dot r^2+\dot z^2)+
l\dot\phi^2,
\end{equation}
where $\epsilon=0,1$ or $-1$ if the geodesics are respectively null,
timelike or spacelike.

\section{Circular geodesics}
Now we restrict ourselves to the study of circular geodesics  assuming
\begin{equation}
\dot r=\dot z=0,\qquad \ddot t=0,\qquad \ddot\phi=0,
\end{equation}
then (31) becomes
\begin{equation}
f^{\prime}\dot t^2-2k^{\prime}\dot t\dot\phi-l^{\prime}\dot\phi^2=0.
\end{equation}
The angular velocity of the test particle is given by
\begin{equation}
\omega=\frac{\dot\phi}{\dot t},
\end{equation}
which becomes using (47)
\begin{equation}
\omega=\frac{1}{l^{\prime}}\left[-k^{\prime}\pm\left(k^{\prime 2}+
f^{\prime}l^{\prime}\right)^{1/2}\right].
\end{equation}
For a stationary spacetime the normal velocity $W^{\mu}$ of the particle
defined as the change in the displacement normal to $\tau^{\mu}=(1,0,0,0)$
relative to its displacement parallel to $\tau^{\mu}$, where $\tau^{\mu}$
is a timelike Killing vector, is \cite{Anderson}
\begin{equation}
W^{\mu}=\left[(-g_{00})^{1/2}\left(dx^0+\frac{g_{0a}}{g_{00}}dx^a\right)
\right]^{-1}V^{\mu},
\end{equation}
where
\begin{equation}
V^{\mu}=\left(-\frac{g_{0a}}{g_{00}}dx^a,dx^1,dx^2,dx^3\right),
\end{equation}
and Latin indexes range from 1 to 3. Considering the metric (1) and the
expression for $\omega$, (48), then (50) becomes
\begin{equation}
W^{\mu}=\frac{(k\omega,0,0,f\omega)}{f^{1/2}(f-k\omega)}.
\end{equation}
The modulus of $W^{\mu}$, defined by $W^2=W^{\mu}W_{\mu}$, and using (52) is
\begin{equation}
W=\frac{D\omega}{f-k\omega}.
\end{equation}
We can write (45) by considering (48)
\begin{equation}
\frac{\epsilon}{\dot t^2}=f-2k\omega-l\omega^2.
\end{equation}
Substituting (53) into (54) we obtain
\begin{equation}
\frac{\epsilon}{\dot t^2}=\frac{D^2f}{(D+kW)^2}(1-W^2),
\end{equation}
which shows that circular geodesics are timelike, null, or spacelike for,
respectively, $W<1,W=1$, or $W>1$.

\section{Circular geodesics in the Weyl class}
In this section we study circular geodesics in the Lewis spacetime for the
Weyl class, which means that the parameters $n,a,b$ and $c$ appearing in
(12,13,14,15) are real.

The parameter $n$ is associated to the Newtonian mass per unit length $\sigma$,
\begin{equation}
\sigma=\frac{1}{4}(1-n),
\end{equation}
of an uniform line mass in the low density regime.
The parameter $a$ is connected to the constant arbitrary potential that
exists in the corresponding Newtonian solution. In the static and locally
flat limit produces angular defficit $\delta$, given by
\begin{equation}
\delta=2\pi\left(1-\frac{1}{a^{1/2}}\right).
\end{equation}
The parameter $b$ is associated, in the locally flat limit, with the
angular momentum of a spinning string. Finally, the parameter $c$ is
related to the vorticity of the source when it is represented by a
stationary completely anisotropic fluid. This parameter, together with the
parameter $b$, is reponsible for the nonstaticity of the metric.
All these interpretations are restricted to the Weyl class only. For
further details see reference \cite{Silva}.

We shall now discuss the expression for the angular velocity (49) and the
tangential velocity (53). In order to exhibit more clearly the role played
by different parameters, we shall consider two different cases, when $b=0,
c\neq0$ and $b\neq, c=0$.

\subsection{Case $b=0, c\neq 0$}
For this case substituting metric (12,13,14,15) into (49) we obtain
\begin{equation}
\omega=\frac{c}{n}\pm\omega_0=\frac{c}{1-4\sigma}\pm\omega_0,
\end{equation}
where $\omega_0$ is the angular velocity when the spacetime is static given
by Levi Civita's metric,
\begin{equation}
\omega_0^2=\frac{1-n}{1+n}a^2r^{-2n}=\frac{2\sigma}{1-2\sigma}a^2
r^{2(4\sigma-1)},
\end{equation}
and (56) has been used.

The term $c/(1-4\sigma)$ represents an inertial frame dragging correction
to the static case, similar to the one appearing in the case of the Kerr
metric \cite{Rindler} or in the field of a massive charged magnetic dipole
\cite{Bonnor}. The presence of $c$ in that term, becomes intelligible when
we recall that this parameter measures the vorticity of the source when
described by a rigidly rotating anisotropic cylinder \cite{Silva}. It
increases or diminishes the modulus of $\omega$ if the vorticity  is in the
same or oposite direction, respectively, of the rotation of the test
particle. Furthermore, the dragging term is increased by the Newtonian mass
per unit length $\sigma$.

Calculating the tangential velocity of the particle (53) with (12,13,14,15)
and (56) we obtain
\begin{eqnarray}
W=\left(\frac{c}{n}\pm \omega_0\right)\frac{r^n}{a}\left(1\pm
\omega_0\frac{cr^{2n}}{na^2}\right)^{-1}\\
=\left[\frac{cr^{1-4\sigma}}{a(1-4\sigma)}\pm W_0\right] \left[1\pm
W_0\frac{cr^{1-4\sigma}}{a(1-4\sigma)}\right]^{-1},
\end{eqnarray}
where $W_0$ is the velocity in the static Levi-Civita spacetime
\begin{equation}
W_0^2=\frac{1-n}{1+n}=\frac{2\sigma}{1-2\sigma},
\end{equation}
which can also be rewritten with (59)
\begin{equation}
W_0=\frac{r^n}{a}\omega_0,
\end{equation}
showing clearly the Newtonian limit $W_0=r\omega_0$.
We see that the dragging term $(cr^{1-4\sigma})/ [a(1-4\sigma)]$ for low
energy densities and $a=1$ is proportional to the vorticity times the
radius. However it is worth noticing that this latter term decreases with
$a$, and furthermore since $W_0<1$, $W$ also decreases with $a$. Thus we
have a dragging effect corrected by a topological deffect. In order to
understand why $W$ decreases with $a$, unlike the angular velocity
$\omega$, which grows with $a$, it should be observed that the arc length
$\Lambda$ along a circular path in this case with $n=1$ becomes
$d\Lambda=(r/a^{1/2})d\phi$. Therefore for a given $d\phi$, $d\Lambda$
decreases with $a$, explaining thereby the decreasing of $W$. On the other
hand, for a given $\Lambda$, $d\phi$ increases with $a$, which explains why
the angular velocity also grows with $a$. This argument is not without its
problems. Indeed if $c=0$ the spacetime reduces to the static Levi-Civita
spacetime but the tangential velocity $W_0$ in unaffec!
ted by $a$ contrary to the argum
ent above. The same is true for the next case below, with respect to $b$.

We notice here too, in this case, that $c$ plays no role as a topological
deffect, i.e., does not appear in the metric term $l$, hence it is
restricted to its frame dragging effect.

Considering (54) and (12,13,14,15) we obtain,
\begin{equation}
\frac{\epsilon}{\dot t^2}=\frac{r^{n+1}}{a}\left[a^2r^{-2n}-\left(\omega
-\frac{c}{n}\right)^2\right],
\end{equation}
and using (56,58,59) it becomes
\begin{equation}
\frac{\epsilon}{\dot t^2}=\frac{r^{n+1}}{a}(a^2r^{-2n}-\omega_0^2)=
\frac{1-4\sigma}{1-2\sigma}ar^{4\sigma}.
\end{equation}
We see from (65) that for circular geodesics the conditions for timelike,
null and spacelike orbits, correspondingly $0\leq\sigma<1/4, \sigma=1/4$
and $1/4<\sigma\leq 1/2$ are unaffected by the vorticity of the source $c$.

\subsection{Case $b\neq 0,c=0$}
For this case substituting the metric (12,13,14,15) into (49) we obtain
\begin{equation}
\omega=\frac{\omega_0(b\omega_0\pm 1)}{1-b^2\omega_0^2}
\end{equation}
Now we have two different kinds of terms appearing in (66). On one hand the
angular velocity corresponding to the Levi-Civita spacetime corrected by
the factor $1-b^2\omega^2_0$, and on the other the topological frame
dragging term associated with $b$. We call it topological frame dragging
since $b$ produces also a topological deffect and to distinguish from the
frame dragging engendered by $c$ that does not produce topological deffect
in the previous case. The increasing of $\omega$ with $b$ may be
interpreted as follows. In the $n=1$ case the line element (1) takes the
form
\begin{equation}
ds^2=-dt^2-2ba^{1/2}dtd\phi+dr^2+dz^2+\left(\frac{r^2}{a}-b^2a\right)
d\phi^2.
\end{equation}
Therefore, from (67), the arc length $\Lambda$ along the circular path is
given by
\begin{equation}
d{\Lambda}=\left(\frac{r^2}{a}-b^2a\right)^{1/2}d\phi.
\end{equation}
Thus, for a given $d\Lambda$, larger $b$'s lead to larger $d\phi$'s and
thereby to larger angular velocities.

Considering (54) and (12,13,14,15) we obtain
\begin{equation}
\frac{\epsilon}{\dot t^2}=ar^{-n+1}(1+b\omega)^2-\frac{r^2}{f}\omega^2,
\end{equation}
and using (56,59,66) it becomes
\begin{equation}
\frac{\epsilon}{\dot t^2}=\frac{1}{(1\mp
b\omega_0)^2}\frac{r^{n+1}}{a}(a^2r^{-2n}-\omega_0^2)=
\frac{1}{(1\mp b\omega_0)^2}\frac{1-4\sigma}{1-2\sigma}ar^{4\sigma}.
\end{equation}
We see from (70) that for circular geodesics the condition for timelike,
null and spacelike orbits are the same as in case $b=0$ and $c\neq 0$.

Calculating the tangential velocity of the particle (53) with (12,13,14,15)
we obtain
\begin{equation}
W=\frac{r^n}{a}\omega_0=W_0.
\end{equation}In the general case, $b\neq 0$ and $c\neq 0$, both parameters
contribute in a cumbersome way to frame dragging and topological effects.

\section{Geodesics in the Weyl class}
Substituting (42,43,44) into (45) we have an expression for the radial
speed ${\dot r}^2$ of the test particle,
\begin {equation}
{\dot r}^2=e^{-\mu}\left(E^2\frac{l}{D^2}+2EL\frac{k}{D^2}-\epsilon-
L^2\frac{f}{D^2}-P^2e^{-\mu}\right).
\end{equation}
We restrict the study of geodesics to $0<n<1$, which is the condition for
circular timelike geodesics as expected in the Newtonian analog
\cite{Silva}.\begin{equation}
{\dot z}=Pr^{(1-n^2)/2},
\end{equation}
which means that, if $P\neq 0$, particles increase their speed along $z$
when distancing radially from de axis, while diminish their axial speed
when moving radially towards the axis. This result indicates that a force
parallel to the $z$ axis appears. In the flat case $n=1$ such effect
vanishes, bringing out its non Newtonian nature.

\begin{equation}
{\ddot z}=\frac{1-n^2}{2}\frac{{\dot r}{\dot z}}{r},
\end{equation}
therefore, this force tends to damp any motion along the $z$ axis whenever
the particle approaches that axis, and reverses this tendency, in the
opposite case.
A similar result is obtained for the van Stockum spacetime \cite{Opher},
although in that case the effect is related to the vorticity of the source,
whereas in our case, neither $b$ nor $c$ appear involved in (73). It is
also worth noticing that non Newtonian forces, parallel to the $z$ axis,
also appear in the field of axially symmetric rotating bodies
\cite{Bonnor1}. However, as in the example of reference \cite{Opher}, the
force parallel to $z$, in \cite{Bonnor1}, unlike our case, is directly
related to the spin of the source. In a stationary rotating universe it has
been observed too a repulsive potential along the axis of rotation
\cite{Schucking}.

Now we study (72) for the two separate cases considered for circular
geodesics and concentrate for null and timelike particles, i.e.
$\epsilon=0$ and $1$.

\subsection{Case $b=0$, $c\neq 0$}
For this case substituting (12,13,14,15) into (72) we obtain
\begin{equation}
{\dot r}^2=r^{-(1-n)^2/2}[V_0-V(r)],
\end{equation}
where
\begin{eqnarray}
V_0=\frac{1}{a}\left(E-L\frac{c}{n}\right)^2>0,\\
V(r)=\epsilon r^{1-n}+P^2r^{(1-n)(3+n)/2}+aL^2r^{-2n}>0.
\end{eqnarray}
The radial acceleration of the particle is from (75)
\begin{eqnarray}
{\ddot r}=\frac{1}{4}r^{-1-(1-n)^2/2}\left[-(1-n)^2V_0-(1-n)(1+n)\epsilon
r^{1-n}\right.\nonumber \\
\left.+(1+n)^2aL^2r^{-2n}-(1-n)(1+n)P^2r^{(1-n)(3+n)/2}\right].
\end{eqnarray}\begin{equation}
V^{\prime}(r)=(1-n)\epsilon r^{-n}-2naL^2r^{-1-2n}+\frac{1}{2}(1-n)(3+n)
P^2r^{-1+(1-n)(3+n)/2}.
\end{equation}
The equation $V^{\prime}(r)=0$ has a solution, say $r=r_c$, which satisfies
\begin{equation}
2naL^2r_c^{-1-2n}=(1-n)\epsilon
r_c^{-n}+\frac{1}{2}(1-n)(3+n)P^2r_c^{-1+(1-n)(3+n)/2}.
\end{equation}
Since
\begin{eqnarray}
V^{\prime\prime}(r)=\frac{1}{r}\left\{-n(1-n)\epsilon r^{-n}+2n(2n+1)aL^2
r^{-1-2n}\right.\nonumber \\
\left.+\frac{1}{2}(1-n)(3+n)\left[\frac{1}{2}(1-n)(3+n)-1\right]P^2
r^{-1+(1-n)(3+n)/2}\right\},
\end{eqnarray}
and with (80) we have
\begin{eqnarray}
V^{\prime\prime}(r_c)=\frac{1}{r_c}\left\{(1-n)(1+n)\epsilon
r_c^{-n}\right. \nonumber\\
\left.+\frac{1}{2}(1-n)(3+n)\left[2n+\frac{1}{2}(1-n)(3+n)\right]P^2
r_c^{-1+(1-n)(3+n)/2}\right\}.
\end{eqnarray}
On the other hand, from (77), if $\epsilon\neq 0$ and/or $P\neq 0$, we have
when $r\rightarrow 0$ or $\infty$, $V(r)\rightarrow\infty$. Furthermore,
from (82) we have $V^{\prime\prime}(r_c)>0$, hence the equation
$V^{\prime}(r)=0$ has only one solution $r=r_c$ and is a minimum.

We see that in this case the parameter $c$ only affects $V_0$, (76), by
modifying the energy of the test particle, and leaving otherwise the
geodesics indistinguishable from the static Levi-Civita spacetime.

\subsubsection{Case $\epsilon =0$, $P=0$}\begin{eqnarray}
V(r)=aL^2r^{-2n},\\
{\ddot r}=\frac{1}{4}r^{-1-(1-n)^2/2}[-(1-n)^2V_0+(1+n)^2aL^2r^{-2n}].
\end{eqnarray}
A null particle with large $r$ approaches $z$ with decreasing negative
acceleration, ${\ddot r}<0$, and increasing speed ${\dot r}$. Its speed
attains a maximum at ${\ddot r}=0$ and then diminishes since the
acceleration becomes positive, ${\ddot r}>0$. The null particle arrives to
a minimum distance $r=r_{min}$ when ${\dot r}=0$ and $V_0=V(r)$. At
$r=r_{min}$ the null particle is reflected to infinity, $r\rightarrow
\infty$, where ${\dot r}\rightarrow 0$.

\subsubsection{Case $\epsilon \neq 0$, $P=0$}\begin{eqnarray}
V(r)=\epsilon r^{1-n}+aL^2r^{-2n},\\
{\ddot r}=\frac{1}{4}r^{-1-(1-n)^2/2}[-(1-n)^2V_0\nonumber\\
-(1-n)(1+n)\epsilon r^{1-n}+(1+n)^2aL^2r^{-2n}].
\end{eqnarray}
For this case $V_0=V(r)$ has two real roots, say $r_{min}$ and $r_{max}$.
The timelike particle approaching $z$ is reflected at $r=r_{min}$ and moves
outwards till it attains ${\dot r}=0$ at $r=r_{max}$ repeating endlessly
this trajectory. This kind of confinement of the test particle has also
been found in the van Stockum spacetime \cite{Opher}. In this case if the
motion is circular, ${\dot r}=0$, there are stable orbits, since $V(r)$ has
a minimum.

\subsubsection{Case $P\neq 0$}
For this case while the particles move along the $z$ axis, as described by
(73), there is a confinement similar to the previous case in the
$z=$constant plane. The effect of $P$, if $\epsilon \neq 0$, is to diminish
the $r_{min}$ and $r_{max}$ attained by the particle. It is interesting to
observe that even if $\epsilon =0$ the null particle has a similar
trajectory in the $z=$constant plane which is not allowed if $P=0$, as
described previously.

\subsection{Case $b\neq 0$, $c=0$}
For this case substituting (12,13,14,15) into (72) we obtain
\begin{equation}
{\dot r}^2=r^{-(1-n)^2/2}[V_0-V(r)],
\end{equation}
where
\begin{eqnarray}
V_0=\frac{1}{a}E^2>0,\\
V(r)=\epsilon r ^{1-n}+P^2r^{(1-n)(3+n)/2}+a(L+bE)^2r^{-2n}>0.
\end{eqnarray}
We see that contrary to $c$, $b$ affects $V(r)$ and not $V_0$, by modifying
the angular momentum of the test particle. From (89) we see that even if
$L=0$ there is an angular momentum term acting on the particle. This
difference between the parameters $b$ and $c$ is expected from our analysis
for the circular geodesics. For $L\neq 0$ and $0$ and $\epsilon =0$, $P=0$;
$\epsilon \neq 0$, $P=0$; and $P\neq 0$ the behaviour of the geodesics is
qualitatively similar to the case $b=0$, $c\neq 0$ with $L\neq 0$.

\end{document}